# Further Results on the Distinctness of Decimations of *l*-sequences

Hong Xu and Wen-Feng Qi

**Abstract**

Let $\underline{a}$ be an *l*-sequence generated by a feedback-with-carry shift register with connection integer $q = p^e$, where $p$ is an odd prime and $e \geq 1$. Goresky and Klapper conjectured that when $p^e \notin \{5, 9, 11, 13\}$, all decimations of $\underline{a}$ are cyclically distinct. When $e = 1$ and $p > 13$, they showed that the set of distinct decimations is large and, in some cases, all deciamtions are distinct. In this article, we further show that when $e \geq 2$ and $p^e \neq 9$, all decimations of $\underline{a}$ are also cyclically distinct.

**Index Terms**

Feedback-with-carry shift register (FCSR), *l*-sequences, arithmetic correlations, 2-adic numbers, integer residue ring, primitive sequences.

## I. INTRODUCTION

Let $p$ be an odd prime and $e \geq 1$ such that 2 is a primitive root modulo $q = p^e$. The class of binary sequences known as *l*-sequences can be described in several ways [7], [8]. An *l*-sequence is the output sequence from a feedback-with-carry shift register (FCSR) with connection integer $q$ whose period reaches the maximum value $\varphi(q)$, where $\varphi$ is Euler's phi function. It is the 2-adic expansions of a rational number $r/q$, where $\gcd(r, q) = 1$. It is also the sequence $a(t) = (A \cdot 2^{-t} (\bmod q))(\bmod 2)$, where $\gcd(A, q) = 1$. These sequences are known to have good statistical properties similar to those of *m*-sequences [7], [8], [9]. They also have the property that the arithmetic correlations between any two cyclically distinct decimations are precisely zero [3].

If $\underline{a} = (a(t))_{t \geq 0}$ is a binary periodic sequence with period $T$, let $\underline{a}^{(d)} = (a(dt))_{t \geq 0}$ denote its $d$-fold decimation and $x^\tau \underline{a} = (a(t+\tau))_{t \geq 0}$ denote the $\tau$-shifted sequence. If $\underline{a}, \underline{b}$ are binary periodic sequences with the same period $T$, we say they are cyclically distinct if $x^\tau \underline{a} \neq \underline{b}$, for every shift $\tau$ with $0 < \tau < T$.

Associate to $\underline{a}$ the formal power series $\alpha = \sum_{t=0}^{\infty} a(t) \cdot 2^t$ and to $x^\tau \underline{b}$ the formal power series $\beta_\tau = \sum_{t=0}^{\infty} b(t+\tau) \cdot 2^t$, which can be regarded as the 2-adic numbers. Let $\gamma = \alpha - \beta_\tau = \sum_{t=0}^{\infty} c(t) \cdot 2^t$ be the difference of the two 2-adic numbers. The sequence of bits $\underline{c} = (c(t))_{t \geq 0}$ is eventually periodic (with period $T$), and the *arithmetic crosscorrelation* $C_{\underline{a},\underline{b}}(\tau)$ is defined to be the number of zeros minus the number of ones in a complete period of length $T$ of $\alpha - \beta_\tau$. The pair of sequences $\underline{a}, \underline{b}$ is said to have *ideal arithmetic cross-correlation* if $C_{\underline{a},\underline{b}}(\tau) = 0$ for every $\tau$ [3].

On the basis of extensive experimental evidence, Goresky and Klapper made the following conjecture.

*Conjecture 1:* [3] Let $\underline{a}$ be an *l*-sequence with connection integer $p^e$ and period $T$. Suppose $p^e \notin \{5, 9, 11, 13\}$, let $c$ and $d$ be relatively prime to $T$ and incongruent modulo $T$. If $\underline{c}$ is the $c$-fold decimation of $\underline{a}$ and $\underline{d}$ is the $d$-fold decimation of $\underline{a}$, then $\underline{c}$ and $\underline{d}$ are cyclically distinct.

This work was supported by NSF of China under Grant number 60373092.

Hong Xu and Wen-Feng Qi are with the department of Applied Mathematics, Zhengzhou Information Engineering University, Zhengzhou, P.R.China. (e-mail: xuhong8096@sina.com and wenfeng.qi@263.net).

Note that the $c$-fold decimation $\underline{c}$ of $\underline{a}$ and the $d$-fold decimation $\underline{d}$ of $\underline{a}$ can be represented as $c(t) = (A \cdot 2^{-ct}(\operatorname{mod} p^e))(\operatorname{mod} 2)$ and $d(t) = (A \cdot 2^{-dt}(\operatorname{mod} p^e))(\operatorname{mod} 2)$, where $2^{-c}(\operatorname{mod} p^e)$ and $2^{-d}(\operatorname{mod} p^e)$ are both primitive roots modulo $p^e$. More generally, let $\xi$ be a primitive root modulo $p^e$, and set $u(t) = A \cdot \xi^t (\operatorname{mod} p^e)$. Then the sequence $\underline{u} = (u(t))_{t \geq 0}$ is a primitive sequence of order 1 over $\mathbb{Z}/(p^e)$ generated by $x - \xi$, and $\underline{u}(\operatorname{mod} 2)$ is an $l$-sequence or its decimation. For the definition of primitive sequences, please see Section II.

For any monic polynomial $f(x)$ over $\mathbb{Z}/(p^e)$, denote $G(f(x), p^e)$ for the set of all sequences over $\mathbb{Z}/(p^e)$ generated by $f(x)$, and set $G'(f(x), p^e) = \{\underline{u} \in G(f(x), p^e) \mid \underline{u} \not\equiv \underline{0}(\operatorname{mod} p)\}$. Detailed description of these two notations, see also Section II. Note that if $\underline{u} \in G(f(x), p^e)$, then $x^k \underline{u} \in G(f(x), p^e)$. With these notations, Conjecture 1 can be restated as follows.

*Conjecture 2:* Let $p^e \notin \{5, 9, 11, 13\}$ with $p$ an odd prime and $e \geq 1$ such that 2 is a primitive root modulo $p^e$. Suppose $\xi$ and $\zeta$ are two different primitive roots modulo $p^e$, and set $f(x) = x - \xi$, $g(x) = x - \zeta$. Then for any $\underline{u} \in G'(f(x), p^e), \underline{v} \in G'(g(x), p^e)$, we have

$$\underline{u} \not\equiv \underline{v}(\operatorname{mod} 2).$$

If this conjecture is proved, then it will provide large families of cyclically distinct sequences with ideal arithmetic correlations. When $e = 1$, some results on the conjecture have been obtained. It's shown in [5] that this conjecture was verified by experiments for all primes $p < 2,000,000$, and asymptotically for large prime $p$, the collection of counterexamples to Conjecture 1 is a vanishingly small fraction of the set of all decimations. Furthermore, it was also shown that if $p = 2r + 1$ and $r = 2s + 1$ with $p, r$, and $s$ prime, then Conjecture 1 holds for $p$.

In this article, we show that Conjecture 1 also holds when $e \geq 2$. The rest of this article is organized as follows. Firstly, an introduction to primitive sequences over integer residue ring and some of their important properties are given in Section II. Next, using the property of primitive sequences, the main result on the distinctness of decimations of $l$-sequences is shown in Section III. Proofs of some lemmas used in Section III are given in Appendix.

Throughout the article, for any positive integers $a$ and $n$, the sign "$a(\operatorname{mod} n)$" refers to the nonnegative minimal residue of $a$ modulo $n$, that is, reducing the number $a$ modulo $n$ to obtain a number between 0 and $n-1$. The notation "$x \equiv a(\operatorname{mod} n)$" is the usual congruent equation, and the notation "$x = a(\operatorname{mod} n)$" means that $x$ is equal to the nonnegative minimal residue of $a(\operatorname{mod} n)$. We also make the covention that whenever a $d$-decimation of a periodic sequence $\underline{a}$ is referred, $d$ is relatively prime to the period of $\underline{a}$.

## II. Prelimilaries

For any odd prime number $p$ and positive integer $e$, let $\mathbb{Z}/(p^e) = \{0, 1, \ldots, p^e - 1\}$ be the integer residue ring modulo $p^e$, and $(\mathbb{Z}/(p^e))^*$ its mutiplicative group. Particularly, $\mathbb{Z}/(p) = \operatorname{GF}(p)$ is the Galois field with $p$ elements.

Let $f(x) = x^n + c_{n-1}x^{n-1} + \cdots + c_0$ be a monic polynomial of degree $n \geq 1$ over $\mathbb{Z}/(p^e)$. If $f(0) \not\equiv 0(\operatorname{mod} p)$, then there exists a positive integer $P$ such that $f(x)$ divides $x^P - 1$ over $\mathbb{Z}/(p^e)$. The least such $P$ is called the period of $f(x)$ over $\mathbb{Z}/(p^e)$ and denoted by $\operatorname{per}(f(x), p^e)$, which is upper bounded by $p^{e-1}(p^n - 1)$ [10]. Moreover, if $\operatorname{per}(f(x), p^e) = p^{e-1}(p^n - 1)$, then say $f(x)$ is a *primitive polynomial* of degree $n$ over $\mathbb{Z}/(p^e)$. In this case, $f(x)(\operatorname{mod} p^i)$ is also a primitive polynomial over $\mathbb{Z}/(p^i)$, whose period is $\text{per}(f(x), p^i) = p^{i-1}(p^n - 1)$, $i = 1, 2, \ldots, e - 1$. Especially, $f(x)(\text{mod } p)$ is a primitive polynomial over the prime field $\text{GF}(p)$.

The sequence $\underline{u} = (u(t))_{t \geq 0}$ over $\mathbb{Z}/(p^e)$ satisfying the recursion

$$u(t + n) = -[c_0 u(t) + c_1 u(t + 1) + \cdots + c_{n-1} u(t + n - 1)](\text{mod } p^e), t \geq 0,$$

is called a *linear recurring sequence* of order $n$ over $\mathbb{Z}/(p^e)$, generated by $f(x)$. Such a sequence is called a *primitive sequence* if $f(x)$ is a primitive polynomial and $\underline{u} \not\equiv \underline{0}(\text{mod } p)$. Particularly, the primitive sequences over $\mathbb{Z}/(p)$ is usually called *m-sequences*.

Denote $G(f(x), p^e)$ for the set of all sequences over $\mathbb{Z}/(p^e)$ generated by $f(x)$, and $G'(f(x), p^e) = \{\underline{u} \in G(f(x), p^e) \mid \underline{u} \not\equiv \underline{0}(\text{mod } p)\}$ for the set of all primitive sequences over $\mathbb{Z}/(p^e)$ generated by $f(x)$.

Any element $v$ in $\mathbb{Z}/(p^e)$ has a unique $p$-adic decomposition as $v = v_0 + v_1 \cdot p + \cdots + v_{e-1} \cdot p^{e-1}$, where $v_i \in \mathbb{Z}/(p)$. Similarly, a sequence $\underline{u}$ over $\mathbb{Z}/(p^e)$ has a unique $p$-adic decomposition as

$$\underline{u} = \underline{u}_0 + \underline{u}_1 \cdot p + \cdots + \underline{u}_{e-1} \cdot p^{e-1},$$

where $\underline{u}_i$ is a sequence over $\mathbb{Z}/(p)$. The sequence $\underline{u}_i$ is called the $i$-th level sequence of $\underline{u}$, and $\underline{u}_{e-1}$ the highest-level sequence of $\underline{u}$. They can be naturally considered as sequences over the prime field $\text{GF}(p)$. Particularly, $\underline{u}_0$ is an $m$-sequence over $\mathbb{Z}/(p)$ generated by $f(x)(\text{mod } p)$ with period $\text{per}(\underline{u}_0) = p^n - 1$.

The following are two important results on primitive polynomials and primitive sequences over $\mathbb{Z}/(p^e)$.

*Proposition 1:* [6] Let $f(x)$ be a primitive polynomial of degree $n$ over $\mathbb{Z}/(p^e)$ with $p$ an odd prime and $e \geq 1$. Then there exists a unique nonzero polynomial $h_f(x)$ over $\mathbb{Z}/(p)$ with $\deg(h_f(x)) < n$, such that

$$x^{p^{i-1}T_0} \equiv 1 + p^i \cdot h_f(x)(\text{mod } f(x), p^{i+1}), i = 1, 2, \ldots, e - 1, \quad (1)$$

where $T_0 = p^n - 1$, the notation "$(\text{mod } f(x), p^{i+1})$" means this congruence equation holds when modulo $f(x)$ and $p^{i+1}$ simultaneously. In other words, we can say $x^{p^{i-1}T_0} \equiv 1 + p^i \cdot h_f(x)(\text{mod } f(x))$ holds over $\mathbb{Z}/(p^{i+1})$ for all $i = 1, 2, \ldots, e - 1$.

*Remark 1:* If $n = 1$, then $h_f(x)$ is a nonzero constant over $\mathbb{Z}/(p)$, denoted by $h_f = h_f(x)$ for simplicity.

*Proposition 2:* [11] Let $f(x)$ be a primitive polynomial of degree $n$ over $\mathbb{Z}/(p^e)$ with $p$ an odd prime and $e \geq 2$. Let $\underline{u} \in G'(f(x), p^e)$, and denote $\underline{\alpha} = h_f(x)\underline{u}_0(\text{mod } p)$, where $h_f(x)$ is defined as (1). Then

$$u_{e-1}(t + j \cdot p^{e-2}T_0) \equiv u_{e-1}(t) + j \cdot \alpha(t)(\text{mod } p), t \geq 0, \quad (2)$$

holds for all $j = 0, 1, \ldots, p - 1$, where $T_0 = p^n - 1$. Furthermore, if $\alpha(t) \neq 0$ for some $t \geq 0$, then

$$\{u_{e-1}(t + j \cdot p^{e-2}T_0) \mid j = 0, 1, \ldots, p - 1\} = \{0, 1, \ldots, p - 1\}. \quad (3)$$

*Remark 2:* Since $\underline{u}_0$ is an $m$-sequence over $\mathbb{Z}/(p)$ generated by $f(x)(\text{mod } p)$ and $\deg(h_f(x)) < \deg(f(x))$, then $\underline{\alpha}$ is also an $m$-sequence over $\mathbb{Z}/(p)$ generated by $f(x)(\text{mod } p)$.

### III. DISINCTNESS OF DECIMATIONS

For convenience, we first review the definition of *l*-sequences.

*Definition 1:* [7][8] An *l*-sequence is a periodic sequence (of period $T = \varphi(q)$) which is obtained from an FCSR with connection integer $q$ for which 2 is a primitive root. Thus $q$ is of the form $q = p^e$, where $p$ is an odd prime and $e \geq 1$.

In this section, we will show that when $e \geq 2$ and $p^e \neq 9$, Conjecture 2 also holds. That is, we have the following theorem.

*Theorem 1:* Let $p^e \neq 9$ with $p$ an odd prime and $e \geq 2$ such that 2 is a primitive root modulo $p^e$. Suppose $\xi$ and $\zeta$ are two different primitive roots modulo $p^e$, and set $f(x) = x - \xi$, $g(x) = x - \zeta$. Then for any $\underline{u} \in G'(f(x), p^e), \underline{v} \in G'(g(x), p^e)$, we have

$$\underline{u} \not\equiv \underline{v} \pmod{2}.$$

Before showing the proof of this theorem, we first give some necessary lemmas. As reference [12] has not yet been published, the proof of the following two lemmas cited from [12] are given in Appendix.

*Lemma 1:* [12] Let $f(x)$ be a primitive polynomial over $\mathbb{Z}/(p)$ with $p$ an odd prime. Then for any $\underline{u}, \underline{v} \in G'(f(x), p)$, $\underline{u} = \underline{v}$ if and only if $\underline{u} \equiv \underline{v} \pmod{2}$.

*Lemma 2:* [12] Let $p$ be an odd prime, $\lambda, \alpha, \beta \in (\mathbb{Z}/(p))^*$ with $\alpha \equiv \lambda\beta \pmod{p}$, and $\delta \in \mathbb{Z}/(p)$ with $\delta \equiv 0 \pmod{2}$. If $1 \leq \lambda \leq p-2$, then there exists a positive integer $j$, $1 \leq j \leq p-1$, such that

$$(j\alpha \pmod{p})(\mod 2) \neq ((j\beta + \delta)(\mod p))(\mod 2).$$

In the following, let $p^e \neq 9$ with $p$ an odd prime and $e \geq 2$ such that 2 is a primitive root modulo $p^e$. Let $\xi$ and $\zeta$ be two different primitive roots modulo $p^e$, and set $f(x) = x - \xi$, $g(x) = x - \zeta$. For any $\underline{u} \in G'(f(x), p^e), \underline{v} \in G'(g(x), p^e)$, we have $\text{per}(\underline{u}) = \text{per}(\underline{v}) = p^{e-1}T_0$, where $T_0 = p - 1$. Let $\underline{\alpha} = h_f\underline{u}_0 \pmod{p}$ and $\underline{\beta} = h_g\underline{v}_0 \pmod{p}$, where $h_f$ and $h_g$ is defined as (1). Since both $\underline{\alpha}$ and $\underline{\beta}$ are $m$-sequences of order 1 generated by $f(x) \pmod{p}$ and $g(x) \pmod{p}$ respectively, then $\text{per}(\underline{\alpha}) = \text{per}(\underline{\beta}) = p - 1$, and $\alpha(t) \neq 0$, $\beta(t) \neq 0$ for all $t \geq 0$.

Similar to the proof in [12], we can show the following two lemmas. Their proofs are also given in Appendix.

*Lemma 3:* Let $\underline{u}, \underline{v}$ be defined as above. If there exists an integer $t$, $t \geq 0$, such that $u_{e-1}(t) \not\equiv v_{e-1}(t) \pmod{2}$, then $\underline{u} \not\equiv \underline{v} \pmod{2}$.

*Lemma 4:* Let $\underline{u}, \underline{v}$, and $\underline{\alpha}, \underline{\beta}$ be defined as above. If $\underline{\alpha} \equiv (p-1)\underline{\beta} \pmod{p}$ and $\underline{u}_{e-1} \equiv \underline{v}_{e-1} \pmod{2}$, then $\underline{u}_{e-1} + \underline{v}_{e-1} \equiv (p-1) \cdot \underline{1} \pmod{p}$.

Next we will show Theorem 1 holds according to $\underline{\alpha} \not\equiv (p-1)\underline{\beta} \pmod{p}$ or $\underline{\alpha} \equiv (p-1)\underline{\beta} \pmod{p}$, respectively.

*Theorem 2:* Let $\underline{u}, \underline{v}$, and $\underline{\alpha}, \underline{\beta}$ be defined as above. If $\underline{\alpha} \not\equiv (p-1)\underline{\beta} \pmod{p}$, then $\underline{u} \not\equiv \underline{v} \pmod{2}$.

*Proof:* By Lemma 3, we need only to show there exists an integer $t$, $t \geq 0$, such that $u_{e-1}(t) \not\equiv v_{e-1}(t) \pmod{2}$.

Since $\underline{\alpha} \not\equiv (p-1)\underline{\beta} \pmod{p}$ and $\text{per}(\underline{\alpha}) = \text{per}(\underline{\beta}) = p-1$, then there exists an integer $t_0$, $0 \leq t_0 \leq p-2$, such that $\alpha(t_0) \not\equiv (p-1)\beta(t_0) \pmod{p}$. From Proposition 2 and (2) we know that

$$u_{e-1}(t_0 + j \cdot p^{e-2}T_0) \equiv u_{e-1}(t_0) + j \cdot \alpha(t_0) \pmod{p},$$

and

$$v_{e-1}(t_0 + j \cdot p^{e-2}T_0) \equiv v_{e-1}(t_0) + j \cdot \beta(t_0) \pmod{p},$$

hold for all $j = 0, 1, \ldots, p-1$, where $T_0 = p - 1$.

On the other hand, as $\alpha(t_0) \neq 0$ and $\beta(t_0) \neq 0$, then by Proposition 2 and (3) we have

$$\{u_{e-1}(t_0 + j \cdot p^{e-2}T_0) | j = 0, 1, \ldots, p-1\} = \{0, 1, \ldots, p-1\},$$

and
$$\{v_{e-1}(t_0 + j \cdot p^{e-2}T_0)|j = 0, 1, ..., p-1\} = \{0, 1, ..., p-1\}.$$

Without loss of generality, let $u_{e-1}(t_0) = 0$, and set $v_{e-1}(t_0) = \delta$.

If $\delta \not\equiv 0 (\mod 2)$, then $u_{e-1}(t_0) \not\equiv v_{e-1}(t_0) (\mod 2)$, and the result holds.

Otherwise, let $\alpha = \alpha(t_0)$ and $\beta = \beta(t_0)$ for simplicity, then $\alpha \neq 0$ and $\beta \neq 0$. Set $\lambda = \alpha\beta^{-1}(\mod p)$, *i.e.*, $\alpha \equiv \lambda\beta(\mod p)$, then $1 \leq \lambda \leq p-2$. From Lemma 2 we know that there exists a positive integer $j_0$, $1 \leq j_0 \leq p-1$, such that

$$(j_0\alpha(\mod p))(\mod 2) \neq ((j_0\beta + \delta)(\mod p))(\mod 2).$$

That is,
$$u_{e-1}(t_0 + j_0 \cdot p^{e-2}T_0) \not\equiv v_{e-1}(t_0 + j_0 \cdot p^{e-2}T_0)(\mod 2).$$

Set $t = t_0 + j_0 \cdot p^{e-2}T_0$, then $u_{e-1}(t) \not\equiv v_{e-1}(t)(\mod 2)$, and the result holds. ∎

*Lemma 5:* Let $f(x), g(x)$ be defined as above. If $p > 3$, there exist no sequences $\underline{u} \in G'(f(x), p^e), \underline{v} \in G'(g(x), p^e)$, such that $\underline{u}_0 = \underline{v}_0$ and $\underline{u}_1 + \underline{v}_1 \equiv (p-1) \cdot \underline{1}(\mod p)$. If $p = 3$ and $e \geq 3$, there exist no sequences $\underline{u} \in G'(f(x), p^e), \underline{v} \in G'(g(x), p^e)$, such that $\underline{u}_0 = \underline{v}_0$ and $\underline{u}_2 + \underline{v}_2 \equiv (p-1) \cdot \underline{1}(\mod p)$.

*Proof:* We first show the case for $p = 3$.

If $e = 3$, it can be shown by experiments that there exist no sequences $\underline{u} \in G'(f(x), 3^3), \underline{v} \in G'(g(x), 3^3)$, such that $\underline{u}_0 = \underline{v}_0$ and $\underline{u}_2 + \underline{v}_2 \equiv (p-1) \cdot \underline{1}(\mod p)$. If $e > 3$, since for any sequences $\underline{u} \in G'(f(x), 3^e), \underline{v} \in G'(g(x), 3^e)$, we have $\underline{u}(\mod 3^3) \in G'(f(x), 3^3)$ and $\underline{v}(\mod 3^3) \in G'(g(x), 3^3)$, thus from above we know that either $\underline{u}_0 \neq \underline{v}_0$ or $\underline{u}_2 + \underline{v}_2 \not\equiv (p-1) \cdot \underline{1}(\mod p)$. So the lemma holds.

Next we show the case for $p > 3$.

Since $\underline{u}_0$ and $\underline{v}_0$ are *m*-sequences over $\mathbb{Z}/(p)$ generated by $f(x)(\mod p)$ and $g(x)(\mod p)$ respectively, then by $\underline{u}_0 = \underline{v}_0$, we have $f(x) \equiv g(x)(\mod p)$, that is, $x - \xi \equiv x - \zeta(\mod p)$. Thus $\xi \equiv \zeta(\mod p)$, so we can set

$$\xi(\mod p^2) = g + k_1p, \text{ and } \zeta(\mod p^2) = g + k_2p,$$

where $g \in (\mathbb{Z}/(p))^*$ is a primitive root modulo $p$, and $k_1, k_2 \in \mathbb{Z}/(p)$.

On the other hand, $\underline{u}(\mod p^2)$ and $\underline{v}(\mod p^2)$ are primitive sequences over $\mathbb{Z}/(p^2)$ generated by $f(x)(\mod p^2)$ and $g(x)(\mod p^2)$ respectively, so we have

$$u(t+1) \equiv u(t) \cdot \xi(\mod p^2), \text{ and } v(t+1) \equiv v(t) \cdot \zeta(\mod p^2).$$

Suppose the lemma does not hold, that is, for all integers $t, t \geq 0$, we have $u_1(t) + v_1(t) \equiv p - 1(\mod p)$. We can derive a contradiction as follows.

Since $\underline{u}_0$ is an *m*-sequence of order 1 over $\mathbb{Z}/(p)$ with $\text{per}(\underline{u}_0) = p-1$, then $\{u_0(t)|t = 0, 1, ..., p-2\} = \{1, 2, ..., p-1\}$. Thus there exist integers $t_1, t_2, 0 \leq t_1, t_2 \leq p-1$, such that $u_0(t_1) = v_0(t_1) = 1$ and $u_0(t_2) = v_0(t_2) = 2$.

Let $w_1 = u_1(t_1), 0 \leq w_1 \leq p-1$. Then by $u_1(t_1) + v_1(t_1) \equiv p - 1(\mod p)$ we have $v_1(t_1) = p - 1 - w_1$. Thus

$$\begin{aligned} u(t_1 + 1) &\equiv u(t_1) \cdot \xi(\mod p^2) \\ &\equiv (1 + w_1p) \cdot (g + k_1p)(\mod p^2) \\ &\equiv g + (gw_1 + k_1) \cdot p(\mod p^2), \end{aligned}$$

and

$$\begin{aligned} v(t_1+1) &\equiv v(t_1)\cdot \zeta \pmod{p^2} \\ &\equiv (1+(p-1-w_1)p)\cdot(g+k_2 p)\pmod{p^2} \\ &\equiv g+(g(p-1-w_1)+k_2)\cdot p\pmod{p^2}. \end{aligned}$$

That is, $u_1(t_1+1) = (gw_1+k_1)\pmod p$, $v_1(t_1+1) = (g(p-1-w_1)+k_2)\pmod p)$. Then by $u_1(t_1+1)+v_1(t_1+1) \equiv p-1 \pmod p$ we have $(gw_1+k_1)+(g(p-1-w_1)+k_2) \equiv p-1 \pmod p$. So we get

$$k_1+k_2 \equiv g-1 \pmod p. \qquad (4)$$

Let $w_2 = u_1(t_2)$, $0 \le w_2 \le p-1$. Then $v_1(t_2) = p-1-w_2$. Similarly we have

$$\begin{aligned} u(t_2+1) &\equiv u(t_2)\cdot \xi \pmod{p^2} \\ &\equiv (2+w_2 p)\cdot(g+k_1 p)\pmod{p^2} \\ &\equiv 2g+(gw_2+2k_1)\cdot p\pmod{p^2}, \end{aligned}$$

and

$$\begin{aligned} v(t_2+1) &\equiv v(t_2)\cdot \zeta \pmod{p^2} \\ &\equiv (2+(p-1-w_2)p)\cdot(g+k_2 p)\pmod{p^2} \\ &\equiv 2g+(g(p-1-w_2)+2k_2)\cdot p\pmod{p^2}. \end{aligned}$$

If $2g < p$, then $u_1(t_2+1) = (gw_2+2k_1)\pmod p$, and $v_1(t_2+1) = (g(p-1-w_2)+2k_2)\pmod p$. Thus by $u_1(t_2+1)+v_1(t_2+1) \equiv p-1 \pmod p$ we have $(gw_2+2k_1)+(g(p-1-w_2)+2k_2) \equiv p-1 \pmod p$, so we get $2(k_1+k_2) \equiv g-1 \pmod p$. Combining with (4), we have $g \equiv 1 \pmod p$, which contradicts the condition that $g$ is a primitive root modulo $p$.

If $2g \ge p$, then $u_1(t_2+1) = (1+gw_2+2k_1)\pmod p$, and $v_1(t_2+1) = (1+g(p-1-w_2)+2k_2)\pmod p$. Thus by $u_1(t_2+1)+v_1(t_2+1) \equiv p-1\pmod p$ we have $(1+gw_2+2k_1)+(1+g(p-1-w_2)+2k_2) \equiv p-1 \pmod p$, so we get $2(k_1+k_2) \equiv g-3 \pmod p$. Combining with (4), we have $g \equiv p-1 \pmod p$, then $g^2 \equiv 1 \pmod p$, which also contradicts the condition that $g$ is a primitive root modulo $p$.

From above analysis we know the assumption that $\underline{u}_1+\underline{v}_1 \equiv (p-1)\cdot\underline{1}\pmod p$ is not correct, thus the lemma holds. ∎

*Theorem 3:* Let $\underline{u},\underline{v}$, and $\underline{\alpha},\underline{\beta}$ be defined as above. If $\underline{\alpha} \equiv (p-1)\underline{\beta}\pmod p$, then $\underline{u} \not\equiv \underline{v}\pmod 2$.

*Proof:* If there exists an integer $t$, $t \ge 0$, such that $u_{e-1}(t) \not\equiv v_{e-1}(t)\pmod 2$, then by Lemma 3, we have $\underline{u} \not\equiv \underline{v}\pmod 2$, and the result holds.

Now suppose $\underline{u}_{e-1} \equiv \underline{v}_{e-1}\pmod 2$. Then by Lemma 4, we have $\underline{u}_{e-1}+\underline{v}_{e-1} \equiv (p-1)\cdot\underline{1}\pmod p$. To show $\underline{u} \not\equiv \underline{v}\pmod 2$, we need to show there exists an integer $t$, such that

$$(\underline{u}(\bmod p^{e-1}))(\bmod 2) \ne (\underline{v}(\bmod p^{e-1}))(\bmod 2).$$

Note that $\underline{u}(\bmod p^{e-1}) = \underline{u}_0+\underline{u}_1\cdot p+\cdots+\underline{u}_{e-2}\cdot p^{e-2}$ and $\underline{v}(\bmod p^{e-1}) = \underline{v}_0+\underline{v}_1\cdot p+\cdots+\underline{v}_{e-2}\cdot p^{e-2}$, which are primitive sequences over $\mathbb{Z}/(p^{e-1})$ generated by $f(x)(\bmod p^{e-1})$ and $g(x)(\bmod p^{e-1})$, respectively. Denote $\underline{u}^{(e-1)} = (\underline{u}(\bmod p^{e-1}))$ and $\underline{v}^{(e-1)} = (\underline{v}(\bmod p^{e-1}))$ for simplicity. If $\underline{u}_{e-2} \not\equiv \underline{v}_{e-2}\pmod 2$, then by Lemma 3 we have $\underline{u}^{(e-1)} \not\equiv \underline{v}^{(e-1)}\pmod 2$, thus $\underline{u} \not\equiv \underline{v}\pmod 2$. So we need only to consider the case when $\underline{u}_{e-2} \equiv \underline{v}_{e-2} (\bmod 2)$. Generally, let $k$ be the largest integer such that $\underline{u}_{e-j} \equiv \underline{v}_{e-j} (\bmod 2)$ for all $1 \leq j \leq k$, $1 \leq k \leq e$.

If $k = e$, then $\underline{u}_j \equiv \underline{v}_j (\bmod 2)$ for all $0 \leq j \leq e-1$. When $j = 0$, we have $\underline{u}_0 \equiv \underline{v}_0 (\bmod 2)$. Since $\underline{u}_0, \underline{v}_0$ are $m$-sequences over $\mathbb{Z}/(p)$ generated by the same primitive polynomial $f(x)(\bmod p) = g(x)(\bmod p)$, then by Lemma 1 we have $\underline{u}_0 = \underline{v}_0$. When $j \geq 1$, then by Lemma 3 we have $\underline{u}_j + \underline{v}_j \equiv (p-1) \cdot \underline{1} (\bmod p)$. This contradicts the result of Lemma 5.

If $k \leq e-1$, then $\underline{u}_{e-k-1} \not\equiv \underline{v}_{e-k-1} (\bmod 2)$. Thus by Lemma 3 we have $\underline{u}^{(e-k)} \not\equiv \underline{v}^{(e-k)} (\bmod 2)$, where $\underline{u}^{(e-k)} = (\underline{u}(\bmod p^{e-k}))$ and $\underline{v}^{(e-k)} = (\underline{v}(\bmod p^{e-k}))$. On the other hand, from the definition of $k$ we know that $\underline{u}_{e-j} \equiv \underline{v}_{e-j} (\bmod 2)$ for all $1 \leq j \leq k$, thus $\underline{u} \not\equiv \underline{v} (\bmod 2)$. So the theorem holds. ∎

*Proof:* [Proof of Theorem 1] If $\underline{\alpha} \not\equiv (p-1)\underline{\beta} (\bmod p)$, then the result holds from Theorem 2. If $\underline{\alpha} \equiv (p-1)\underline{\beta} (\bmod p)$, then the result holds from Theorem 3. ∎

*Remark 3:* When $e \geq 1$ and $p > 13$, Goresky and Klapper showed that almost all decimations of $l$-sequences with prime connection integer $p$ are cyclically distinct [5]. In this article, we further show that when $e \geq 2$ and $p^e \neq 9$, all decimations of $l$-sequences with connection integer $p^e$ are also cyclically distinct, which completes the proof of Conjecture 1.

## APPENDIX

In this section, we give the proof of some lemmas used in Section III. The first two lemmas are results cited from [12], and the other two lemmas can be proved using similar method as in [12]. Here we include their proofs for completeness.

*Lemma 1:* [12] Let $f(x)$ be a primitive polynomial over $Z/(p)$ with $p$ an odd prime. Then for any $\underline{u}, \underline{v} \in G'(f(x), p)$, $\underline{u} = \underline{v}$ if and only if $\underline{u} \equiv \underline{v} (\mod 2)$.

*Proof:* The necessary condition is obvious. We need only to show if $\underline{u} \equiv \underline{v}(\mod 2)$, then $\underline{u} = \underline{v}$.

If $\underline{u}$ and $\underline{v}$ are linear dependent over $\mathbb{Z}/(p)$, that is, there exists an integer $\lambda \in (\mathbb{Z}/(p))^*$, such that $\underline{v} \equiv \lambda \cdot \underline{u}(\mod p)$. If $\lambda$ is even, let $t$ be an integer such that $u(t) = 1$, then $u(t) \not\equiv v(t)(\mod 2)$, which is in contradiction with $\underline{u} \equiv \underline{v}(\mod 2)$. If $\lambda$ is odd and $\lambda \neq 1$, let $k$ be the least positive integer such that $(k-1)\lambda < p < k\lambda$, and let $t$ be an integer such that $u(t) = k$. Since $(k\lambda(\mod p))(\mod 2) = (k\lambda - p)(\mod 2) \neq k(\mod 2)$, then $u(t) \not\equiv v(t)(\mod 2)$, which is also in contradiction with $\underline{u} \equiv \underline{v}(\mod 2)$. Thus $\lambda = 1$ and $\underline{u} = \underline{v}$.

If $\underline{u}$ and $\underline{v}$ are linear independent over $\mathbb{Z}/(p)$, since $\underline{u}$ and $\underline{v}$ are $m$-sequences generated by the same polynomial $f(x)$, then there exists an integer $t$ such that $u(t) = 0$ and $v(t) = 1$. So we have $u(t) \not\equiv v(t)(\mod 2)$, which is also in contradiction with $\underline{u} \equiv \underline{v}(\mod 2)$. Thus $\underline{u} = \underline{v}$. ∎

*Remark 4:* In this article, $\deg(f(x)) = 1$, and $\underline{u}, \underline{v}$ are linear dependent over $Z/(p)$.

*Lemma 2:* [12] Let $p$ be an odd prime, $\lambda, \alpha, \beta \in (Z/(p))^*$ with $\alpha \equiv \lambda\beta(\mod p)$, and $\delta \in Z/(p)$ with $\delta \equiv 0(\mod 2)$. If $1 \leq \lambda \leq p-2$, then there exists a positive integer $j$, $1 \leq j \leq p-1$, such that

$$(j\alpha(\mod p))(\mod 2) \neq ((j\beta + \delta)(\mod p))(\mod 2).$$

*Proof:* Since $\alpha \equiv \lambda\beta(\mod p)$, we have

$$\{(j \cdot \alpha(\mod p), (j \cdot \beta + \delta)(\mod p)) \mid j = 0, 1, \ldots, p-1\}$$
$$= \{(j \cdot \lambda(\mod p), (j + \delta)(\mod p)) \mid j = 0, 1, \ldots, p-1\}.$$

Thus we need only to show there exists a positive integer $j$, $1 \leq j \leq p-1$, such that

$$(j\lambda(\mod p))(\mod 2) \neq ((j + \delta)(\mod p))(\mod 2). \quad (5)$$

1. $\lambda = 1$.

As $\delta$ is even, set $j = p - \delta$, then $j\lambda(\mod p) = p - \delta$ is odd, but $(j+\delta)(\mod p) = 0$ is even, thus (5) holds.

2. $2 \leq \lambda \leq p - 2$, and $\delta < p - 1$.

If $\lambda$ is even, set $j = 1$, then $j\lambda(\mod p) = \lambda$ is even, but $(j+\delta)(\mod p) = 1+\delta$ is odd, thus (5) holds.

If $\lambda$ is odd, let $k_1$ be the least positive integer such that $(k_1 - 1)\lambda < p < k_1\lambda < 2p$ and $k_2$ be the least positive integer such that $(k_2 - 1)\lambda < 2p < k_2\lambda < 3p$. It's clear that $2 \leq k_1 < k_2 < p$.

(1.1) If $k_1 < p - \delta$, then $(k_1\lambda(\mod p))(\mod 2) = (k_1\lambda - p)(\mod 2) \neq k_1(\mod 2)$, but $(k_1 + \delta)(\mod 2) = k_1(\mod 2)$. Set $j = k_1$, then (5) holds.

(1.2) If $k_1 = p - \delta$, from $k_2 > k_1 = p - \delta$ and $2p < k_2\lambda < 3p$ we have $(k_2\lambda(\mod p))(\mod 2) = k_2\lambda - 2p(\mod 2) = k_2(\mod 2)$, and $((k_2 + \delta)(\mod p))(\mod 2) = (k_2 + \delta - p)(\mod 2) \neq k_2(\mod 2)$. Set $j = k_2$, then (5) holds.

(1.3) If $k_1 > p - \delta$, then from the definition of $k_1$ we know that $0 < (p - \delta)\lambda < p$. Set $j = p - \delta$, then $j\lambda(\mod p) = (p - \delta)\lambda$ is odd, but $(j + \delta)(\mod p) = 0$ is even, thus (5) holds.

3. $2 \leq \lambda \leq p - 2$, and $\delta = p - 1$.

In this case, we need only to show there exists a positive integer $j$, $1 \leq j \leq p - 1$, such that

$$(j\lambda(\mod p))(\mod 2) \neq ((j - 1)(\mod p))(\mod 2). \tag{6}$$

If $\lambda$ is odd, set $j = 1$, then $j\lambda(\mod p) = \lambda$ is odd, but $(j - 1)(\mod p) = 0$ is even, thus (6) holds.

If $\lambda$ is even, then $2 \leq \lambda \leq p - 3$. Let $k$ be the least positive integer such that $k\lambda(\mod p) < p - \lambda$. As $2 \leq \lambda \leq p-3$, then $p - \lambda \geq 3$, and $1 \leq k \leq p-2$. Thus $(k\lambda(\mod p))(\mod 2) = ((k+1)\lambda(\mod p))(\mod 2)$. Since $(k - 1)(\mod 2) \neq k(\mod 2)$, set $j_1 = k$, $j_2 = k + 1$, then either $(j_1\lambda(\mod p))(\mod 2) \neq ((j_1 - 1)(\mod p))(\mod 2)$ or $(j_2\lambda(\mod p))(\mod 2) \neq ((j_2 - 1)(\mod p))(\mod 2)$, thus (6) holds. ∎

*Remark 5:* The case when $\delta = 0$ is not included in the original result of [12], but the proof is the same, so we include here.

With the same notations as Section 3, in the following, let $p^e \neq 9$ with $p$ an odd prime number and $e \geq 2$ such that 2 is a primitive root modulo $p^e$. Let $\xi$ and $\zeta$ be two different primitive roots modulo $p^e$, and set $f(x) = x - \xi, g(x) = x - \zeta$. For any $\underline{u} \in G'(f(x), p^e), \underline{v} \in G'(g(x), p^e)$, let $\underline{\alpha} = h_f\underline{u}_0(\mod p)$ and $\underline{\beta} = h_g\underline{v}_0(\mod p)$, where $h_f$ and $h_g$ is defined as (1). In this case, $\alpha(t) \neq 0$, $\beta(t) \neq 0$ hold for all integers $t \geq 0$.

*Lemma 3:* Let $\underline{u}, \underline{v}$ be defined as above. If there exists an integer $t, t \geq 0$, such that $u_{e-1}(t) \not\equiv v_{e-1}(t)(\mod 2)$, then $\underline{u} \not\equiv \underline{v}(\mod 2)$.

*Proof:* Since for all integers $t \geq 0$, $\alpha(t) \neq 0$ and $\beta(t) \neq 0$, then by Proposition 2 and (3), we have

$$\{u_{e-1}(t + j \cdot p^{e-2}T_0)|j = 0, 1, ..., p - 1\} = \{0, 1, ..., p - 1\}, t \geq 0, \tag{7}$$

and

$$\{v_{e-1}(t + j \cdot p^{e-2}T_0)|j = 0, 1, ..., p - 1\} = \{0, 1, ..., p - 1\}, t \geq 0, \tag{8}$$

where $T_0 = p - 1$. Thus for any fixed integer $t \geq 0$, from (7) and (8) we know that $u_{e-1}(t + j \cdot p^{e-2}T_0)$ and $v_{e-1}(t + j \cdot p^{e-2}T_0)$ belong to the same set $\{0, 1, ..., p - 1\}$, whose cardinality $p$ is odd.

If there exists an integer $t_0, t_0 \geq 0$, such that $u_{e-1}(t_0) \not\equiv v_{e-1}(t_0)(\mod 2)$, then from above we know that there also exists an integer $j_0$, $1 \leq j_0 \leq p - 1$, such that

$$u_{e-1}(t_0 + j_0 \cdot p^{e-2}T_0) \equiv v_{e-1}(t_0 + j_0 \cdot p^{e-2}T_0)(\mod 2).$$

On the other hand, for $e \geq 2$ we have

$$\underline{u} = \underline{u}(\mod p^{e-1}) + \underline{u}_{e-1} \cdot p^{e-1}, \text{ and } \text{per}(\underline{u}(\mod p^{e-1})) = p^{e-2}(p - 1),$$

and

$$\underline{v} = \underline{v}(\mod p^{e-1}) + \underline{v}_{e-1} \cdot p^{e-1}, \text{ and } \text{per}(\underline{v}(\mod p^{e-1})) = p^{e-2}(p - 1).$$

Thus the two congruence equations $u(t_0) \equiv v(t_0)(\mod 2)$ and $u(t_0+j_0 \cdot p^{e-2}T_0) \equiv v(t_0+j_0 \cdot p^{e-2}T_0)(\mod 2)$ can not hold simultaneously. That is, there exists an integer $t$, such that

$$u(t) \not\equiv v(t)(\mod 2).$$

So we get
$$\underline{u} \not\equiv \underline{v} (\mathrm{mod}\, 2).$$

∎

*Lemma 4:* Let $\underline{u}, \underline{v}$, and $\underline{\alpha}, \underline{\beta}$ be defined as above. If $\underline{\alpha} \equiv (p-1)\underline{\beta}(\mathrm{mod}\, p)$ and $\underline{u}_{e-1} \equiv \underline{v}_{e-1}(\mathrm{mod}\, 2)$, then $\underline{u}_{e-1} + \underline{v}_{e-1} \equiv (p-1) \cdot \underline{1}(\mathrm{mod}\, p)$.

*Proof:* Set $T_0 = p-1$. From Proposition 2 and (2) we know that
$$u_{e-1}(t + j \cdot p^{e-2}T_0) \equiv u_{e-1}(t) + j \cdot \alpha(t)(\mathrm{mod}\, p),\ t \geq 0, \tag{9}$$
and
$$v_{e-1}(t + j \cdot p^{e-2}T_0) \equiv v_{e-1}(t) + j \cdot \beta(t)(\mathrm{mod}\, p),\ t \geq 0, \tag{10}$$
holds for all $j = 0, 1, \ldots, p-1$.

Since $\underline{\alpha} \equiv (p-1)\underline{\beta}(\mathrm{mod}\, p)$, that is, $\underline{\alpha} + \underline{\beta} \equiv \underline{0}(\mathrm{mod}\, p)$, thus for all integers $t \geq 0$, $\alpha(t) + \beta(t) \equiv 0(\mathrm{mod}\, p)$ holds. Combining with (9) and (10), we have
$$u_{e-1}(t + j \cdot p^{e-2}T_0) + v_{e-1}(t + j \cdot p^{e-2}T_0) \equiv u_{e-1}(t) + v_{e-1}(t)(\mathrm{mod}\, p),$$
for all integers $t \geq 0$ and $j = 0, 1, \ldots, p-1$.

For any fixed $t \geq 0$, let $\tau = (u_{e-1}(t) + v_{e-1}(t))(\mathrm{mod}\, p)$, then $u_{e-1}(t + j \cdot p^{e-2}T_0) + v_{e-1}(t + j \cdot p^{e-2}T_0) \equiv \tau(\mathrm{mod}\, p)$ for all $j = 0, 1, \ldots, p-1$. Next we will show $\tau = p-1$.

As $\alpha(t) \neq 0$, then by Proposition 2 and (3) we have
$$\{u_{e-1}(t + j \cdot p^{e-2}T_0)|j = 0, 1, ..., p-1\} = \{0, 1, ..., p-1\}.$$
Thus there exist integers $j_0, j_1$, $0 \leq j_0, j_1 \leq p-1$, such that $u_{e-1}(t + j_0 \cdot p^{e-2}T_0) = 0$ and $u_{e-1}(t + j_1 \cdot p^{e-2}T_0) = p-1$.

If $\tau$ is odd, then from $u_{e-1}(t+j_0 \cdot p^{e-2}T_0) = 0$ and $u_{e-1}(t+j_0 \cdot p^{e-2}T_0) + v_{e-1}(t+j_0 \cdot p^{e-2}T_0) \equiv \tau(\mathrm{mod}\, p)$ we know that $v_{e-1}(t + j_0 \cdot p^{e-2}T_0) = \tau$ is odd, thus $u_{e-1}(t + j_0 \cdot p^{e-2}T_0) \not\equiv v_{e-1}(t + j_0 \cdot p^{e-2}T_0)(\mathrm{mod}\, 2)$. This contradicts the condition that $\underline{u}_{e-1} \equiv \underline{v}_{e-1}(\mathrm{mod}\, 2)$.

If $\tau$ is even and $\tau \neq p-1$, then from $u_{e-1}(t+j_1 \cdot p^{e-2}T_0) = p-1$ and $u_{e-1}(t+j_1 \cdot p^{e-2}T_0) + v_{e-1}(t+j_1 \cdot p^{e-2}T_0) \equiv \tau(\mathrm{mod}\, p)$ we know that $v_{e-1}(t+j_1 \cdot p^{e-2}T_0) = \tau+1$ is odd, thus $u_{e-1}(t+j_1 \cdot p^{e-2}T_0) \not\equiv v_{e-1}(t+j_1 \cdot p^{e-2}T_0)(\mathrm{mod}\, 2)$. This also contradicts the condition that $\underline{u}_{e-1} \equiv \underline{v}_{e-1}(\mathrm{mod}\, 2)$.

Therefore $\tau = p-1$, that is, $u_{e-1}(t) + v_{e-1}(t) \equiv p-1(\mathrm{mod}\, p)$ for all integers $t \geq 0$, so the lemma holds. ∎